\documentclass[12pt]{article}
\usepackage{a4,epsfig,amssymb}
\def\PL#1#2#3{{ Phys. Lett. }{ B#1 }(#2) #3}
\def\NIM#1#2#3{{ Nucl. Inst. Meth. }{ #1 }(#2) #3}
\def\ra{\rightarrow} 
\newcommand{\PW}{\mbox{$\mathrm{W}$}}
\newcommand{\PWm}{\mbox{$\mathrm{W}^-$}}

\newcommand{\PZz}{\mbox{$\mathrm{Z}$}}
\newcommand{\PMW}{\mbox{$m_{\mathrm{W}}$}}
\newcommand{\PMMW}{\mbox{$m_{\mathrm{W}}^2$}}
\newcommand{\PMMZ}{\mbox{$m_{\mathrm{Z}}^2$}}
\newcommand{\PSW}{\mbox{$\sigma_{\mathrm{WW}}$}}
\newcommand{\PSWv}{\mbox{$\sigma_{\mathrm{WW}}^{\mathrm{vis}}$}}
\newcommand{\PGW}{\mbox{$\Gamma_{\mathrm{W}}$}}
\newcommand{\PGWi}{\mbox{$\Gamma_{\mathrm{W}}^{\mathrm{invis}}$}}
\newcommand{\PGWv}{\mbox{$\Gamma_{\mathrm{W}}^{\mathrm{vis}}$}}
\newcommand{\GEV}{\mbox{$\mathrm{GeV}$}}

\newcommand{\GEVcc}{\mbox{$\mathrm{GeV}/{{\it c}^2}$}}
\newcommand{\MEVc}{\mbox{$\mathrm{MeV}/{\it c}$}}
\newcommand{\MEVcc}{\mbox{$\mathrm{MeV}/{{\it c}^2}$}}
\newcommand{\Pem}{\mbox{$\mathrm{e}^-$}}

\newcommand{\Pepem}{\mbox{$\mathrm{e}^+\mathrm{e}^-$}}
\newcommand{\Pee}{\mbox{$\mathrm{ee}$}}
\newcommand{\PWpWm}{\mbox{$\mathrm{W}^+\mathrm{W}^-$}}
\newcommand{\PWW}{\mbox{$\mathrm{W}\mathrm{W}$}}
\newcommand{\PZZ}{\mbox{$\mathrm{Z}\mathrm{Z}$}}

\newcommand{\ckm}[1]{\ensuremath{V_\mathrm{#1}}}
\newcommand{\ev}{\mbox{$\mathrm{e}\nu$}}
\newcommand{\mv}{\mbox{$\mu\nu$}}
\newcommand{\tv}{\mbox{$\tau\nu$}}
\newcommand{\lv}{\mbox{$\ell\nu$}}
\newcommand{\qq}{\mbox{$\mathrm{qq}$}}
\newcommand{\qqbar}{\mbox{$\mathrm{q\bar{q}}$}}
\newcommand{\stat}{\ensuremath{\mathrm { (stat.)}}}
\newcommand{\syst}{\ensuremath{\mathrm { (syst.)}}}
\newcommand{\pb}{\ensuremath{\mathrm {pb}}}
\newcommand{\ipb}{\ensuremath{\mathrm {pb}^{-1}}}

\begin{document}

\begin{titlepage}

{\Large\centerline{EUROPEAN LABORATORY FOR PARTICLE PHYSICS (CERN)}}
\vspace{1cm}
\begin{flushright}
CERN-EP/99-035 \\
1st March 1999 \\
\end{flushright}   

\begin{center}
\vspace{3cm}
\boldmath
{\huge\bf Measurement of W-pair
production in  \Pepem{} collisions
at~183 GeV}\\
{\Large \vspace{7ex} The ALEPH Collaboration}
\unboldmath
\end{center}

\vspace{1cm}
\begin{center}
{\bf Abstract}
\end{center}
The production of $\PWpWm$ pairs
is analysed in a data sample collected by ALEPH at  
a mean centre-of-mass energy of 182.7~\GEV, 
corresponding to an integrated
luminosity of 57~$\ipb$.
Cross sections are given for 
different topologies of W decays
into leptons or hadrons.
Under Standard Model assumptions for 
the W-pair production and decay,
the  W-pair cross section is measured to be
$
15.57 \pm 0.62 \stat  \pm 0.29 \syst ~\pb.
$
Using also the W-pair data samples collected by ALEPH at 
lower centre-of-mass energies, the decay branching ratio
of the W boson into
hadrons is measured to be~$\mathrm{B (W\ra hadrons)} = 
68.93 \pm 1.21 \stat  
\pm 0.51 \syst\%$,
allowing a determination of
the CKM matrix element~$|\ckm{cs}|= 1.043 \pm 0.058\stat \pm 0.026\syst$.
The agreement of the cross sections with the Standard Model prediction 
allows a limit to be set on the W decay rate to 
undetectable final states.
\\[3cm]

\centerline{\em (to be Submitted to Physics Letters B)}
\end{titlepage}

\newpage

\pagestyle{empty}
\newpage
\small
%
%
\newlength{\saveparskip}
\newlength{\savetextheight}
\newlength{\savetopmargin}
\newlength{\savetextwidth}
\newlength{\saveoddsidemargin}
\newlength{\savetopsep}
\setlength{\saveparskip}{\parskip}
\setlength{\savetextheight}{\textheight}
\setlength{\savetopmargin}{\topmargin}
\setlength{\savetextwidth}{\textwidth}
\setlength{\saveoddsidemargin}{\oddsidemargin}
\setlength{\savetopsep}{\topsep}
%
%
\setlength{\parskip}{0.0cm}
\setlength{\textheight}{25.0cm}
\setlength{\topmargin}{-1.5cm}
\setlength{\textwidth}{16 cm}
\setlength{\oddsidemargin}{-0.0cm}
\setlength{\topsep}{1mm}
\pretolerance=10000
\centerline{\large\bf The ALEPH Collaboration}
\footnotesize
\vspace{0.5cm}
{\raggedbottom
\begin{sloppypar}
\samepage\noindent
R.~Barate,
D.~Decamp,
P.~Ghez,
C.~Goy,
S.~Jezequel,
J.-P.~Lees,
F.~Martin,
E.~Merle,
\mbox{M.-N.~Minard},
B.~Pietrzyk
\nopagebreak
\begin{center}
\parbox{15.5cm}{\sl\samepage
Laboratoire de Physique des Particules (LAPP), IN$^{2}$P$^{3}$-CNRS,
F-74019 Annecy-le-Vieux Cedex, France}
\end{center}\end{sloppypar}
\vspace{2mm}
\begin{sloppypar}
\noindent
R.~Alemany,
M.P.~Casado,
M.~Chmeissani,
J.M.~Crespo,
E.~Fernandez,
M.~Fernandez-Bosman,
Ll.~Garrido,$^{15}$
E.~Graug\`{e}s,
A.~Juste,
M.~Martinez,
G.~Merino,
R.~Miquel,
Ll.M.~Mir,
P.~Morawitz,
A.~Pacheco,
I.C.~Park,
I.~Riu
\nopagebreak
\begin{center}
\parbox{15.5cm}{\sl\samepage
Institut de F\'{i}sica d'Altes Energies, Universitat Aut\`{o}noma
de Barcelona, 08193 Bellaterra (Barcelona), E-Spain$^{7}$}
\end{center}\end{sloppypar}
\vspace{2mm}
\begin{sloppypar}
\noindent
A.~Colaleo,
D.~Creanza,
M.~de~Palma,
G.~Gelao,
G.~Iaselli,
G.~Maggi,
M.~Maggi,
S.~Nuzzo,
A.~Ranieri,
G.~Raso,
F.~Ruggieri,
G.~Selvaggi,
L.~Silvestris,
P.~Tempesta,
A.~Tricomi,$^{3}$
G.~Zito
\nopagebreak
\begin{center}
\parbox{15.5cm}{\sl\samepage
Dipartimento di Fisica, INFN Sezione di Bari, I-70126 Bari, Italy}
\end{center}\end{sloppypar}
\vspace{2mm}
\begin{sloppypar}
\noindent
X.~Huang,
J.~Lin,
Q. Ouyang,
T.~Wang,
Y.~Xie,
R.~Xu,
S.~Xue,
J.~Zhang,
L.~Zhang,
W.~Zhao
\nopagebreak
\begin{center}
\parbox{15.5cm}{\sl\samepage
Institute of High-Energy Physics, Academia Sinica, Beijing, The People's
Republic of China$^{8}$}
\end{center}\end{sloppypar}
\vspace{2mm}
\begin{sloppypar}
\noindent
D.~Abbaneo,
U.~Becker,$^{19}$
G.~Boix,$^{6}$
M.~Cattaneo,
V.~Ciulli,
G.~Dissertori,
H.~Drevermann,
R.W.~Forty,
M.~Frank,
F.~Gianotti,
A.W.~Halley,
J.B.~Hansen,
J.~Harvey,
P.~Janot,
B.~Jost,
I.~Lehraus,
O.~Leroy,
C.~Loomis,
P.~Maley,
P.~Mato,
A.~Minten,
A.~Moutoussi,
F.~Ranjard,
L.~Rolandi,
D.~Rousseau,
D.~Schlatter,
M.~Schmitt,$^{20}$
O.~Schneider,$^{23}$
W.~Tejessy,
F.~Teubert,
I.R.~Tomalin,
E.~Tournefier,
M.~Vreeswijk,
A.E.~Wright
\nopagebreak
\begin{center}
\parbox{15.5cm}{\sl\samepage
European Laboratory for Particle Physics (CERN), CH-1211 Geneva 23,
Switzerland}
\end{center}\end{sloppypar}
\vspace{2mm}
\begin{sloppypar}
\noindent
Z.~Ajaltouni,
F.~Badaud
G.~Chazelle,
O.~Deschamps,
S.~Dessagne,
A.~Falvard,
C.~Ferdi,
P.~Gay,
C.~Guicheney,
P.~Henrard,
J.~Jousset,
B.~Michel,
S.~Monteil,
\mbox{J-C.~Montret},
D.~Pallin,
P.~Perret,
F.~Podlyski
\nopagebreak
\begin{center}
\parbox{15.5cm}{\sl\samepage
Laboratoire de Physique Corpusculaire, Universit\'e Blaise Pascal,
IN$^{2}$P$^{3}$-CNRS, Clermont-Ferrand, F-63177 Aubi\`{e}re, France}
\end{center}\end{sloppypar}
\vspace{2mm}
\begin{sloppypar}
\noindent
J.D.~Hansen,
J.R.~Hansen,
P.H.~Hansen,
B.S.~Nilsson,
B.~Rensch,
A.~W\"a\"an\"anen
\nopagebreak
\begin{center}
\parbox{15.5cm}{\sl\samepage
Niels Bohr Institute, 2100 Copenhagen, DK-Denmark$^{9}$}
\end{center}\end{sloppypar}
\vspace{2mm}
\begin{sloppypar}
\noindent
G.~Daskalakis,
A.~Kyriakis,
C.~Markou,
E.~Simopoulou,
A.~Vayaki
\nopagebreak
\begin{center}
\parbox{15.5cm}{\sl\samepage
Nuclear Research Center Demokritos (NRCD), GR-15310 Attiki, Greece}
\end{center}\end{sloppypar}
\vspace{2mm}
\begin{sloppypar}
\noindent
A.~Blondel,
\mbox{J.-C.~Brient},
F.~Machefert,
A.~Roug\'{e},
M.~Swynghedauw,
R.~Tanaka,
A.~Valassi,$^{12}$
H.~Videau
\nopagebreak
\begin{center}
\parbox{15.5cm}{\sl\samepage
Laboratoire de Physique Nucl\'eaire et des Hautes Energies, Ecole
Polytechnique, IN$^{2}$P$^{3}$-CNRS, \mbox{F-91128} Palaiseau Cedex, France}
\end{center}\end{sloppypar}
\vspace{2mm}
\begin{sloppypar}
\noindent
E.~Focardi,
G.~Parrini,
K.~Zachariadou
\nopagebreak
\begin{center}
\parbox{15.5cm}{\sl\samepage
Dipartimento di Fisica, Universit\`a di Firenze, INFN Sezione di Firenze,
I-50125 Firenze, Italy}
\end{center}\end{sloppypar}
\vspace{2mm}
\begin{sloppypar}
\noindent
R.~Cavanaugh,
M.~Corden,
C.~Georgiopoulos
\nopagebreak
\begin{center}
\parbox{15.5cm}{\sl\samepage
Supercomputer Computations Research Institute,
Florida State University,
Tallahassee, FL 32306-4052, USA $^{13,14}$}
\end{center}\end{sloppypar}
\vspace{2mm}
\begin{sloppypar}
\noindent
A.~Antonelli,
G.~Bencivenni,
G.~Bologna,$^{4}$
F.~Bossi,
P.~Campana,
G.~Capon,
F.~Cerutti,
V.~Chiarella,
P.~Laurelli,
G.~Mannocchi,$^{5}$
F.~Murtas,
G.P.~Murtas,
L.~Passalacqua,
M.~Pepe-Altarelli$^{1}$
\nopagebreak
\begin{center}
\parbox{15.5cm}{\sl\samepage
Laboratori Nazionali dell'INFN (LNF-INFN), I-00044 Frascati, Italy}
\end{center}\end{sloppypar}
\vspace{2mm}
\begin{sloppypar}
\noindent
M.~Chalmers,
L.~Curtis,
J.G.~Lynch,
P.~Negus,
V.~O'Shea,
B.~Raeven,
C.~Raine,
D.~Smith,
P.~Teixeira-Dias,
A.S.~Thompson,
J.J.~Ward
\nopagebreak
\begin{center}
\parbox{15.5cm}{\sl\samepage
Department of Physics and Astronomy, University of Glasgow, Glasgow G12
8QQ,United Kingdom$^{10}$}
\end{center}\end{sloppypar}
\begin{sloppypar}
\noindent
O.~Buchm\"uller,
S.~Dhamotharan,
C.~Geweniger,
P.~Hanke,
G.~Hansper,
V.~Hepp,
E.E.~Kluge,
A.~Putzer,
J.~Sommer,
K.~Tittel,
S.~Werner,$^{19}$
M.~Wunsch
\nopagebreak
\begin{center}
\parbox{15.5cm}{\sl\samepage
Institut f\"ur Hochenergiephysik, Universit\"at Heidelberg, D-69120
Heidelberg, Germany$^{16}$}
\end{center}\end{sloppypar}
\vspace{2mm}
\begin{sloppypar}
\noindent
R.~Beuselinck,
D.M.~Binnie,
W.~Cameron,
P.J.~Dornan,$^{1}$
M.~Girone,
S.~Goodsir,
N.~Marinelli,
E.B.~Martin,
J.~Nash,
J.~Nowell,
A.~Sciab\`a,
J.K.~Sedgbeer,
P.~Spagnolo,
E.~Thomson,
M.D.~Williams
\nopagebreak
\begin{center}
\parbox{15.5cm}{\sl\samepage
Department of Physics, Imperial College, London SW7 2BZ,
United Kingdom$^{10}$}
\end{center}\end{sloppypar}
\vspace{2mm}
\begin{sloppypar}
\noindent
V.M.~Ghete,
P.~Girtler,
E.~Kneringer,
D.~Kuhn,
G.~Rudolph
\nopagebreak
\begin{center}
\parbox{15.5cm}{\sl\samepage
Institut f\"ur Experimentalphysik, Universit\"at Innsbruck, A-6020
Innsbruck, Austria$^{18}$}
\end{center}\end{sloppypar}
\vspace{2mm}
\begin{sloppypar}
\noindent
A.P.~Betteridge,
C.K.~Bowdery,
P.G.~Buck,
P.~Colrain,
G.~Crawford,
G.~Ellis,
A.J.~Finch,
F.~Foster,
G.~Hughes,
R.W.L.~Jones,
N.A.~Robertson,
M.I.~Williams
\nopagebreak
\begin{center}
\parbox{15.5cm}{\sl\samepage
Department of Physics, University of Lancaster, Lancaster LA1 4YB,
United Kingdom$^{10}$}
\end{center}\end{sloppypar}
\vspace{2mm}
\begin{sloppypar}
\noindent
P.~van~Gemmeren,
I.~Giehl,
F.~H\"olldorfer,
C.~Hoffmann,
K.~Jakobs,
K.~Kleinknecht,
M.~Kr\"ocker,
H.-A.~N\"urnberger,
G.~Quast,
B.~Renk,
E.~Rohne,
H.-G.~Sander,
S.~Schmeling,
H.~Wachsmuth
C.~Zeitnitz,
T.~Ziegler
\nopagebreak
\begin{center}
\parbox{15.5cm}{\sl\samepage
Institut f\"ur Physik, Universit\"at Mainz, D-55099 Mainz, Germany$^{16}$}
\end{center}\end{sloppypar}
\vspace{2mm}
\begin{sloppypar}
\noindent
J.J.~Aubert,
C.~Benchouk,
A.~Bonissent,
J.~Carr,$^{1}$
P.~Coyle,
A.~Ealet,
D.~Fouchez,
F.~Motsch,
P.~Payre,
M.~Talby,
M.~Thulasidas,
A.~Tilquin
\nopagebreak
\begin{center}
\parbox{15.5cm}{\sl\samepage
Centre de Physique des Particules, Facult\'e des Sciences de Luminy,
IN$^{2}$P$^{3}$-CNRS, F-13288 Marseille, France}
\end{center}\end{sloppypar}
\vspace{2mm}
\begin{sloppypar}
\noindent
M.~Aleppo,
M.~Antonelli,
F.~Ragusa
\nopagebreak
\begin{center}
\parbox{15.5cm}{\sl\samepage
Dipartimento di Fisica, Universit\`a di Milano e INFN Sezione di
Milano, I-20133 Milano, Italy.}
\end{center}\end{sloppypar}
\vspace{2mm}
\begin{sloppypar}
\noindent
R.~Berlich,
V.~B\"uscher,
H.~Dietl,
G.~Ganis,
K.~H\"uttmann,
G.~L\"utjens,
C.~Mannert,
W.~M\"anner,
\mbox{H.-G.~Moser},
S.~Schael,
R.~Settles,
H.~Seywerd,
H.~Stenzel,
W.~Wiedenmann,
G.~Wolf
\nopagebreak
\begin{center}
\parbox{15.5cm}{\sl\samepage
Max-Planck-Institut f\"ur Physik, Werner-Heisenberg-Institut,
D-80805 M\"unchen, Germany\footnotemark[16]}
\end{center}\end{sloppypar}
\vspace{2mm}
\begin{sloppypar}
\noindent
P.~Azzurri,
J.~Boucrot,
O.~Callot,
S.~Chen,
M.~Davier,
L.~Duflot,
\mbox{J.-F.~Grivaz},
Ph.~Heusse,
A.~Jacholkowska,
M.~Kado,
J.~Lefran\c{c}ois,
L.~Serin,
\mbox{J.-J.~Veillet},
I.~Videau,$^{1}$
J.-B.~de~Vivie~de~R\'egie,
D.~Zerwas
\nopagebreak
\begin{center}
\parbox{15.5cm}{\sl\samepage
Laboratoire de l'Acc\'el\'erateur Lin\'eaire, Universit\'e de Paris-Sud,
IN$^{2}$P$^{3}$-CNRS, F-91898 Orsay Cedex, France}
\end{center}\end{sloppypar}
\vspace{2mm}
\begin{sloppypar}
\noindent
G.~Bagliesi,
S.~Bettarini,
T.~Boccali,
C.~Bozzi,$^{24}$
G.~Calderini,
R.~Dell'Orso,
I.~Ferrante,
A.~Giassi,
A.~Gregorio,
F.~Ligabue,
A.~Lusiani,
P.S.~Marrocchesi,
A.~Messineo,
F.~Palla,
G.~Rizzo,
G.~Sanguinetti,
G.~Sguazzoni,
R.~Tenchini,
C.~Vannini,
A.~Venturi,
P.G.~Verdini
\samepage
\begin{center}
\parbox{15.5cm}{\sl\samepage
Dipartimento di Fisica dell'Universit\`a, INFN Sezione di Pisa,
e Scuola Normale Superiore, I-56010 Pisa, Italy}
\end{center}\end{sloppypar}
\vspace{2mm}
\begin{sloppypar}
\noindent
G.A.~Blair,
J.~Coles,
G.~Cowan,
M.G.~Green,
D.E.~Hutchcroft,
L.T.~Jones,
T.~Medcalf,
J.A.~Strong,
J.H.~von~Wimmersperg-Toeller
\nopagebreak
\begin{center}
\parbox{15.5cm}{\sl\samepage
Department of Physics, Royal Holloway \& Bedford New College,
University of London, Surrey TW20 OEX, United Kingdom$^{10}$}
\end{center}\end{sloppypar}
\vspace{2mm}
\begin{sloppypar}
\noindent
D.R.~Botterill,
R.W.~Clifft,
T.R.~Edgecock,
P.R.~Norton,
J.C.~Thompson
\nopagebreak
\begin{center}
\parbox{15.5cm}{\sl\samepage
Particle Physics Dept., Rutherford Appleton Laboratory,
Chilton, Didcot, Oxon OX11 OQX, United Kingdom$^{10}$}
\end{center}\end{sloppypar}
\vspace{2mm}
\begin{sloppypar}
\noindent
\mbox{B.~Bloch-Devaux},
P.~Colas,
B.~Fabbro,
G.~Fa\"if,
E.~Lan\c{c}on,
\mbox{M.-C.~Lemaire},
E.~Locci,
P.~Perez,
H.~Przysiezniak,
J.~Rander,
\mbox{J.-F.~Renardy},
A.~Rosowsky,
A.~Trabelsi,$^{21}$
B.~Tuchming,
B.~Vallage
\nopagebreak
\begin{center}
\parbox{15.5cm}{\sl\samepage
CEA, DAPNIA/Service de Physique des Particules,
CE-Saclay, F-91191 Gif-sur-Yvette Cedex, France$^{17}$}
\end{center}\end{sloppypar}
\vspace{2mm}
\begin{sloppypar}
\noindent
S.N.~Black,
J.H.~Dann,
H.Y.~Kim,
N.~Konstantinidis,
A.M.~Litke,
M.A. McNeil,
G.~Taylor
\nopagebreak
\begin{center}
\parbox{15.5cm}{\sl\samepage
Institute for Particle Physics, University of California at
Santa Cruz, Santa Cruz, CA 95064, USA$^{22}$}
\end{center}\end{sloppypar}
\vspace{2mm}
\begin{sloppypar}
\noindent
C.N.~Booth,
S.~Cartwright,
F.~Combley,
P.N.~Hodgson,
M.S.~Kelly,
M.~Lehto,
L.F.~Thompson
\nopagebreak
\begin{center}
\parbox{15.5cm}{\sl\samepage
Department of Physics, University of Sheffield, Sheffield S3 7RH,
United Kingdom$^{10}$}
\end{center}\end{sloppypar}
\vspace{2mm}
\begin{sloppypar}
\noindent
K.~Affholderbach,
A.~B\"ohrer,
S.~Brandt,
C.~Grupen,
A.~Misiejuk,
G.~Prange,
U.~Sieler
\nopagebreak
\begin{center}
\parbox{15.5cm}{\sl\samepage
Fachbereich Physik, Universit\"at Siegen, D-57068 Siegen, Germany$^{16}$}
\end{center}\end{sloppypar}
\vspace{2mm}
\begin{sloppypar}
\noindent
G.~Giannini,
B.~Gobbo
\nopagebreak
\begin{center}
\parbox{15.5cm}{\sl\samepage
Dipartimento di Fisica, Universit\`a di Trieste e INFN Sezione di Trieste,
I-34127 Trieste, Italy}
\end{center}\end{sloppypar}
\vspace{2mm}
\begin{sloppypar}
\noindent
J.~Putz,
J.~Rothberg,
S.~Wasserbaech,
R.W.~Williams
\nopagebreak
\begin{center}
\parbox{15.5cm}{\sl\samepage
Experimental Elementary Particle Physics, University of Washington, WA 98195
Seattle, U.S.A.}
\end{center}\end{sloppypar}
\vspace{2mm}
\begin{sloppypar}
\noindent
S.R.~Armstrong,
E.~Charles,
P.~Elmer,
D.P.S.~Ferguson,
Y.~Gao,
S.~Gonz\'{a}lez,
T.C.~Greening,
O.J.~Hayes,
H.~Hu,
S.~Jin,
P.A.~McNamara III,
J.M.~Nachtman,$^{2}$
J.~Nielsen,
W.~Orejudos,
Y.B.~Pan,
Y.~Saadi,
I.J.~Scott,
J.~Walsh,
Sau~Lan~Wu,
X.~Wu,
G.~Zobernig
\nopagebreak
\begin{center}
\parbox{15.5cm}{\sl\samepage
Department of Physics, University of Wisconsin, Madison, WI 53706,
USA$^{11}$}
\end{center}\end{sloppypar}
}
\footnotetext[1]{Also at CERN, 1211 Geneva 23, Switzerland.}
\footnotetext[2]{Now at University of California at Los Angeles (UCLA),
Los Angeles, CA 90024, U.S.A.}
\footnotetext[3]{Also at Dipartimento di Fisica, INFN Sezione di Catania,
95129 Catania, Italy.}
\footnotetext[4]{Also Istituto di Fisica Generale, Universit\`{a} di
Torino, 10125 Torino, Italy.}
\footnotetext[5]{Also Istituto di Cosmo-Geofisica del C.N.R., Torino,
Italy.}
\footnotetext[6]{Supported by the Commission of the European Communities,
contract ERBFMBICT982894.}
\footnotetext[7]{Supported by CICYT, Spain.}
\footnotetext[8]{Supported by the National Science Foundation of China.}
\footnotetext[9]{Supported by the Danish Natural Science Research Council.}
\footnotetext[10]{Supported by the UK Particle Physics and Astronomy Research
Council.}
\footnotetext[11]{Supported by the US Department of Energy, grant
DE-FG0295-ER40896.}
\footnotetext[12]{Now at LAL, 91898 Orsay, France.}
\footnotetext[13]{Supported by the US Department of Energy, contract
DE-FG05-92ER40742.}
\footnotetext[14]{Supported by the US Department of Energy, contract
DE-FC05-85ER250000.}
\footnotetext[15]{Permanent address: Universitat de Barcelona, 08208 Barcelona,
Spain.}
\footnotetext[16]{Supported by the Bundesministerium f\"ur Bildung,
Wissenschaft, Forschung und Technologie, Germany.}
\footnotetext[17]{Supported by the Direction des Sciences de la
Mati\`ere, C.E.A.}
\footnotetext[18]{Supported by Fonds zur F\"orderung der wissenschaftlichen
Forschung, Austria.}
\footnotetext[19]{Now at SAP AG, 69185 Walldorf, Germany}
\footnotetext[20]{Now at Harvard University, Cambridge, MA 02138, U.S.A.}
\footnotetext[21]{Now at D\'epartement de Physique, Facult\'e des Sciences de Tunis, 1060 Le Belv\'ed\`ere, Tunisia.}
\footnotetext[22]{Supported by the US Department of Energy,
grant DE-FG03-92ER40689.}
\footnotetext[23]{Now at Universit\'e de Lausanne, 1015 Lausanne, Switzerland.}
\footnotetext[24]{Now at INFN Sezione di Ferrara, 44100 Ferrara, Italy.}
%
%
\setlength{\parskip}{\saveparskip}
\setlength{\textheight}{\savetextheight}
\setlength{\topmargin}{\savetopmargin}
\setlength{\textwidth}{\savetextwidth}
\setlength{\oddsidemargin}{\saveoddsidemargin}
\setlength{\topsep}{\savetopsep}
\normalsize
\newpage
\pagestyle{plain}
\setcounter{page}{1}

\section{Introduction}
This letter presents results on  W-pair production
using data collected with the ALEPH detector at 
centre-of-mass energies ranging from 180.8 to 183.8 \GEV, 
during the 1997 data taking period.

The experimental conditions and data analysis follow closely 
those used in the cross section measurement at lower LEP2 
energies.
As they are already described in detail in~\cite{xsecA},
attention will be focused on changes in selection procedures
other than a simple
rescaling of cuts with the increased collision energy.

At these energies
the cross section has a much smaller dependence 
on the W mass with respect to the threshold productions, and 
the total production rate constitutes a test of the Standard Model (SM).

About 900 W pairs are expected
in the collected data sample, allowing improved direct determinations 
of the W hadronic and leptonic branching ratios.
The hadronic branching ratio is sensitive to the yet poorly known
coupling $|\ckm{cs}|$ of the W to cs pairs.

A detailed description of the ALEPH detector can be found in
Ref.~\cite{det}
and of its performance in Ref.~\cite{perf}. 
The  luminosity is measured with small-angle Bhabha events,  
using lead-proportional wire sampling calorimeters~\cite{ALEPHlcal},
with an accepted Bhabha cross section of approximately 4.6 nb~\cite{BHLUMI}.
An integrated luminosity of 
56.81~$\pm$~0.11~$\stat~\pm$~0.29~$\syst~\ipb$
was recorded,
at a mean centre-of-mass energy of 182.66~$\pm$~0.05~\GEV~\cite{LEPbeam}.

\section{Physics processes and definition of the W-pair cross section}
\label{sec:xsdef}

To lowest order within the Standard Model, three diagrams contribute to
W-pair production in \Pepem annihilations:  the $s$-channel $\gamma$ and
Z boson exchange and the $t$-channel $\nu_{\mathrm{e}}$ exchange, 
referred to as CC03 diagrams.
Each W is expected to decay rapidly into a quark-antiquark pair 
($\PWm\ra\overline{\mathrm u}{\mathrm d}, \overline{\mathrm c}{\mathrm d}
,\overline{\mathrm u}{\mathrm s}, \overline{\mathrm c}{\mathrm s},
\overline{\mathrm u}{\mathrm b}, \overline{\mathrm c}{\mathrm b} $)
or a lepton-neutrino pair 
($\PWm\ra \Pem\overline{\nu}_{\mathrm{e}}, 
\mu^-\overline{\nu}_{\mu},\tau^-\overline{\nu}_{\tau}$).
Therefore, each CC03 diagram leads to an experimentally accessible 
four-fermion final state.
However, many other Standard Model processes can lead to the same 
four-fermion final states as W-pair decays, interfering with the CC03
diagrams.

In the following, 
all signal cross sections are defined 
by the production of four-fermion final states only through two resonating 
W bosons, and will be referred as CC03 cross sections. 
The effect of the non-CC03 diagrams is corrected for with additive
terms, obtained by comparing 
Monte Carlo (MC) simulations including only CC03 processes with all
Standard Model four-fermion final states 
compatible with W-pair decays (WW-like four-fermion final states), 
following the same procedure as in Ref.~\cite{xsecA}.
Such Monte Carlo additive terms are referred to in the following as 
``4f-CC03 corrections''. 
They are computed separately for all different
WW-like four-fermion final states $f_1 f_2 f_3 f_4$, as
$(\epsilon_{\rm 4f}\sigma_{\rm 4f}-
  \epsilon_{\rm CC03}\sigma_{\rm CC03})^{f_1 f_2 f_3 f_4}$,
where $\epsilon$ and $\sigma$ refer to MC selection efficiencies and generator 
cross sections, on samples of $f_1 f_2 f_3 f_4$ final states generated 
with the full four-fermion productions (4f subscript) or 
with W-pair decays only (CC03 subscript).
All the signal efficiencies quoted in the following are to be 
interpreted as CC03 efficiencies, evaluated on such Monte Carlo 
samples.

Two Monte Carlo event generators are used to simulate the signal events.
Samples of events are generated with different W masses, 
both for CC03 diagrams and for  
all WW-like four-fermion diagrams, with
{\tt KORALW}~\cite{KORALW}.
A comparison sample is generated 
with {\tt EXCALIBUR}~\cite{EXCALIBUR} with and without colour reconnection
effects (following the ansatz of Ref.~\cite{COLOUR_RECONNECTION}).
The {\tt KORALW} samples serve to determine the 
four-fermion and CC03 efficiencies used 
to obtain the final result. 
Other {\tt KORALW} samples are used to check 
the \PMW~dependence of 
the selection procedures and of the four-fermion to CC03 correction. 
The {\tt EXCALIBUR} samples 
are used as a check of 
the Monte Carlo simulation of the physics processes, 
and  to assess the effects of colour reconnection.

The {\tt PYTHIA 5.7}~\cite{PYTHIA} Monte Carlo program
is used to generate background events coming from \qqbar,
\PZZ, \PZz\Pee ~and \PW\ev ~processes.
Other background events are generated with 
{\tt PHOT02}~\cite{PHOT02} for gamma-gamma interactions,
{\tt KORALZ}~\cite{KORALZ} for $\mu$ and $\tau$ pair production,
{\tt BHWIDE}~\cite{BHWIDE} and {\tt UNIBAB}~\cite{UNIBAB} 
for Bhabha scattering events.
For the fully leptonic selections, 
non-WW-like four-fermion background processes with
two visible leptons (as $\ell'\ell'\nu_{\ell}\nu_{\ell}$)
are generated with {\tt EXCALIBUR}.
Care is taken to avoid double counting of backgrounds with
different generators and to avoid counting as backgrounds  
those WW-like four-fermion processes 
that are part of the 4f-CC03 correction. 
Monte Carlo samples corresponding to integrated luminosities
at least twenty times as large as that of the data
are fully simulated for all background reactions.

\boldmath
\section{Selection of W-pair candidates}
\subsection{$ \PWW \ra \ell\nu \ell\nu$ events}
\unboldmath

The selection of fully leptonic W-pair decays is an update
of the two selections used for the cross section and 
branching ratio measurements at 161 and 172~GeV~\cite{xsecA},
to the 183 GeV energy regime.

In the first selection, 
which does not make use of lepton identification criteria,
two changes are introduced.
The lepton-lepton acoplanarity is required to be smaller than 
$175^{\circ}$  and
the energy of the leading lepton
must be smaller than 86~GeV. 
In the data, 47 events are selected with this analysis.

In the second selection
lepton identification is used to classify events 
in six different di-lepton channels 
(ee, e$\mu$, e$\tau$, $\mu\mu$, $\mu\tau$ or $\tau\tau$),
according to the flavour of leptons,
and optimised cuts are applied individually in each channel.
The electron and muon identification criteria are the same as those 
used for the leptonic classifications at lower energies~\cite{xsecA}.
A lepton then is classified as a tau 
either if no lepton identification is fulfilled 
or if the identified lepton has an energy lower 
than 25 GeV (0.137$\sqrt{s}$).
In the data, 57 events are selected with this second analysis.

The inclusive combination of the two selections 
has an efficiency of 71.5\%
for the fully leptonic WW channels, combined assuming lepton universality.
For the cross sections and the branching ratio measurements,
the events selected by the combination of the two analyses
are classified into six channels according to
the di-lepton flavour classification used for the second selection.
The CC03 efficiencies in the individual $\ell\nu\ell\nu$ channels
are given in Table~\ref{tab:summ}
for the inclusive combination of the two selections.
The total background amounts to 151 fb
and is dominated by $\gamma\gamma \ra \ell\ell$ 
and non-WW-like four-fermion events, 
as $\Pepem\!\ra \ell'\ell'\nu_{\ell}\nu_{\ell}$
with $\ell'\neq\ell $.
In the data, the inclusive combination selects 61 events.

\renewcommand{\arraystretch}{1.2}
\begin{table}[htbp]
 \begin{center}
\caption{\sl Summary of results of the different event selections
  on Monte Carlo and Data events. Efficiencies are given in percent of 
  CC03~processes. The \tv\qq~selection column refers to the performance of the 
  exclusive selection. The \qq\qq~column refers to events with a NN 
  output greater than -0.2 . 
  The listed backgrounds do not include the 4f-CC03 corrections. 
  In the \qq\qq~column the backgrounds include non-qqqq WW decays.}
\vspace{6pt}
\label{tab:summ}
 \begin{tabular}{|c|l|cccccc|ccc|c||c|} \hline
\multicolumn{2}{|c|}{} &\multicolumn{11}{c|}{Event selection and classification}    \\  
 \cline{3-13}
\multicolumn{2}{|c|}{} & 
 $\ev\ev\!\!$&$\ev\mv\!\!$&$\ev\tv\!\!$&$\mv\mv\!\!$&$\mv\tv\!\!$&
 $\tv\tv\!\!$&$\ev\qq\!\!\!$&$\mv\qq\!\!\!$&$\tv\qq\!\!\!$&$\qq\qq\!\!\!\!$& All\\ 
 \hline
& $\ev\ev\!\!$  &68.2 &   -  &  8.4 &   -  &   -  &   -  &   -  &   -  &   -  &   -  & 76.6 \\
& $\ev\mv\!\!$  & 0.1 & 70.8 &  2.0 &   -  &  2.9 &  0.3 &   -  &   -  &   -  &   -  & 76.1 \\
& $\ev\tv\!\!$  & 4.3 &  3.8 & 57.9 &   -  &  0.1 &  3.5 &  0.3 &   -  &   -  &   -  & 69.9 \\
Eff. for
& $\mv\mv\!\!$  &  -  &   -  &   -  & 71.1 &  4.7 &  0.2 &   -  &   -  &   -  &   -  & 76.0 \\
$\PWW\!\!\ra$
& $\mv\tv\!\!$  &  -  &  4.1 &  0.2 &  3.9 & 61.7 &  1.5 &   -  &  0.6 &   -  &   -  & 72.0 \\
(in \%)
& $\tv\tv\!\!$  &  -  &  0.7 &  5.1 &  0.3 &  5.6 & 45.2 &   -  &   -  &   -  &   -  & 56.9 \\ 
 \cline{2-13}
& $\ev\qq\!\!\!$  &  -  &   -  &   -  &   -  &   -  &   -  & 81.2 &  0.3 &  6.2 &   -  & 87.7 \\ 
& $\mv\qq\!\!\!$  &  -  &   -  &   -  &   -  &   -  &   -  &  0.2 & 88.6 &  3.5 &   -  & 92.3 \\
& $\tv\qq\!\!\!$  &  -  &   -  &   -  &   -  &   -  &   -  &  3.1 &  3.3 & 54.1 &   -  & 60.5 \\ 
 \cline{2-13}
& $\qq\qq\!\!\!\!$  &  -  &   -  &   -  &   -  &   -  &   -  &  0.1 &   -  &  0.1 & 84.3 & 84.5 \\ 
 \hline\hline
\multicolumn{2}{|c|}{\parbox{2.5cm}{Backgrounds (in pb)}} &
  0.02 & 0.01 & 0.05 &  0.01 & 0.02 & 0.05 & 0.11& 0.05 & 0.11 &1.24 &1.67 \\
 \hline\hline
\multicolumn{2}{|c|}{\parbox[c]{2.5cm}{Observed Events}} &
   6 &  14 &  18 &   8 &  11 &   4 & 127 & 113 &  86 & 432 & 818 \\
 \hline
\end{tabular}
\end{center}
\end{table}
\renewcommand{\arraystretch}{1.0}

\begin{table}[htb]
\begin{center}
\caption[]{\sl Systematic errors summary in fb.}
\vspace{6pt}
\label{tabsys}
\begin{tabular}{|l|c|c|c|}
\cline{2-4}
\multicolumn{1}{c|}{} & \multicolumn{3}{|c|}{WW cross section}  \\
\hline
Source &  $\ell\nu\ell\nu$ &
$\ell\nu\qq $ & $\qq\qq$ \\
\hline
Calibration of calorimeters  & & &  10 \\
\hline
Preselection and Shape variables  & & & 155 \\
\hline
WW generator and  $\PMW $ dependence & & & 105 \\ 
\hline
WW fragmentation    & & &  60 \\ 
\hline
QCD generator      & & & 60 \\
\hline
Final State Interactions & & & 40 \\ 
\hline
Background normalisation & 17 & 80 & 120 \\
\hline
Luminosity & 9 & 40 & 50 \\
\hline
MC statistics & 28 & 54 & 20 \\
\hline
Photon veto   & 23 & 70 & \\
\hline
Lepton id    & 5 &  30 &  \\
\hline
Lepton isolation  & & 40  &\\ 
\hline
Probability discrimination & & 40 & \\
\hline
\hline
Total     & 41 &  141 &  247 \\
\hline
\end{tabular}
\end{center}
\end{table}

Beam related background and detector noise 
are not fully reproduced by the simulation
and affect the efficiency of the photon veto cuts.
Random trigger events
are used to compute the inefficiency introduced.
The result is 
(4.3$\pm$0.8)\% for the low angle photon veto and 
is applied as a correction factor lowering the 
Monte Carlo efficiencies.
The systematic error takes into account
the uncertainty on the inefficiencies.
For the large angle photon veto 
these effects are not corrected for,
but a systematic error of $\pm1.7\%$ is assigned to them,
again from the study of random trigger events.
Lepton identification efficiency has a 2\% systematic uncertainty,
to cover residual discrepancies in identification efficiency and purity
between LEP1 data and Monte Carlo.
The final sources of systematic errors 
are listed Table~\ref{tabsys}.
The effect on the total cross section amounts to 0.04 pb 
and is dominated by the Monte Carlo statistical errors 
in the subtraction of backgrounds
and in the signal efficiencies.

A likelihood fit is applied to determine
individual cross sections for each fully leptonic decay channel
using efficiency matrices for CC03 
and backgrounds, as in Table~\ref{tab:summ}.
WW-like four-fermion processes are taken into account
in the 4f-CC03 correction
for each individual channel, given by
$(\epsilon_{\rm 4f}\sigma_{\rm 4f}-
  \epsilon_{\rm CC03}\sigma_{\rm CC03})^{\lv\lv}$. 
For all channels together, 
this correction amounts to $+6\pm 19$ fb,
where the uncertainty comes from Monte Carlo statistics.
 
The results of the fit, where the cross sections are constrained 
to be positive, yield
\begin{eqnarray*}
\sigma(\PWW\ra \ev\ev ) & = &
 0.12 ^{+0.08}_{-0.06} \stat 
\pm 0.01 \syst ~\pb, \\
\sigma(\PWW\ra \mv\mv ) & = &
 0.17 ^{+0.08}_{-0.07} \stat 
\pm 0.01 \syst ~\pb, \\
\sigma(\PWW\ra \tv\tv ) & = &
 0.00 ^{+0.08}_{-0.00} \stat 
\pm 0.03 \syst ~\pb, \\
\sigma(\PWW\ra \ev\mv ) & = &
 0.33 ^{+0.11}_{-0.09} \stat 
\pm 0.02 \syst ~\pb, \\
\sigma(\PWW\ra \ev\tv ) & = &
 0.44 ^{+0.15}_{-0.12} \stat 
\pm 0.03 \syst ~\pb, \\
\sigma(\PWW\ra \mv\tv ) & = &
 0.27 ^{+0.11}_{-0.09} \stat 
\pm 0.02 \syst ~\pb,
\end{eqnarray*}
where the statistical errors correspond to a variation of the likelihood
logarithm of 0.5 around the fitted maximum.
The systematic errors are obtained by varying all input
parameters in the fit
according to their uncertainties.
For $\sigma(\PWW\ra \tau\nu\tau\nu )$, 
the systematic uncertainty
refers to the positive statistical error, rather than to the 
fitted central value.

The fully leptonic cross section is obtained with the same fit, 
assuming lepton universality, and yields
$$
\sigma (\PWW\ra \ell\nu\ell\nu) 
= 1.38 \pm 0.20 \stat \pm 0.04 \syst ~\pb.
$$


\boldmath
\subsection{$\PWW \ra \ell \nu \qq$ events}
\unboldmath

The three  selection procedures developed for the lower energy 
measurements are applied.
One selection is optimised for WW events with electrons or muons and
requires an energetic identified electron or muon, while
the other  two  are designed
for $\tau\nu \qq$ events, based on 
global variables and topological properties of events.

\boldmath
\subsubsection{$\PWW\ra \mathrm{e}\nu \qq$ and 
$\PWW \ra \mu \nu \qq$ selections}
\unboldmath

\begin{figure}[htb]
  \centerline{
    \epsfig{file=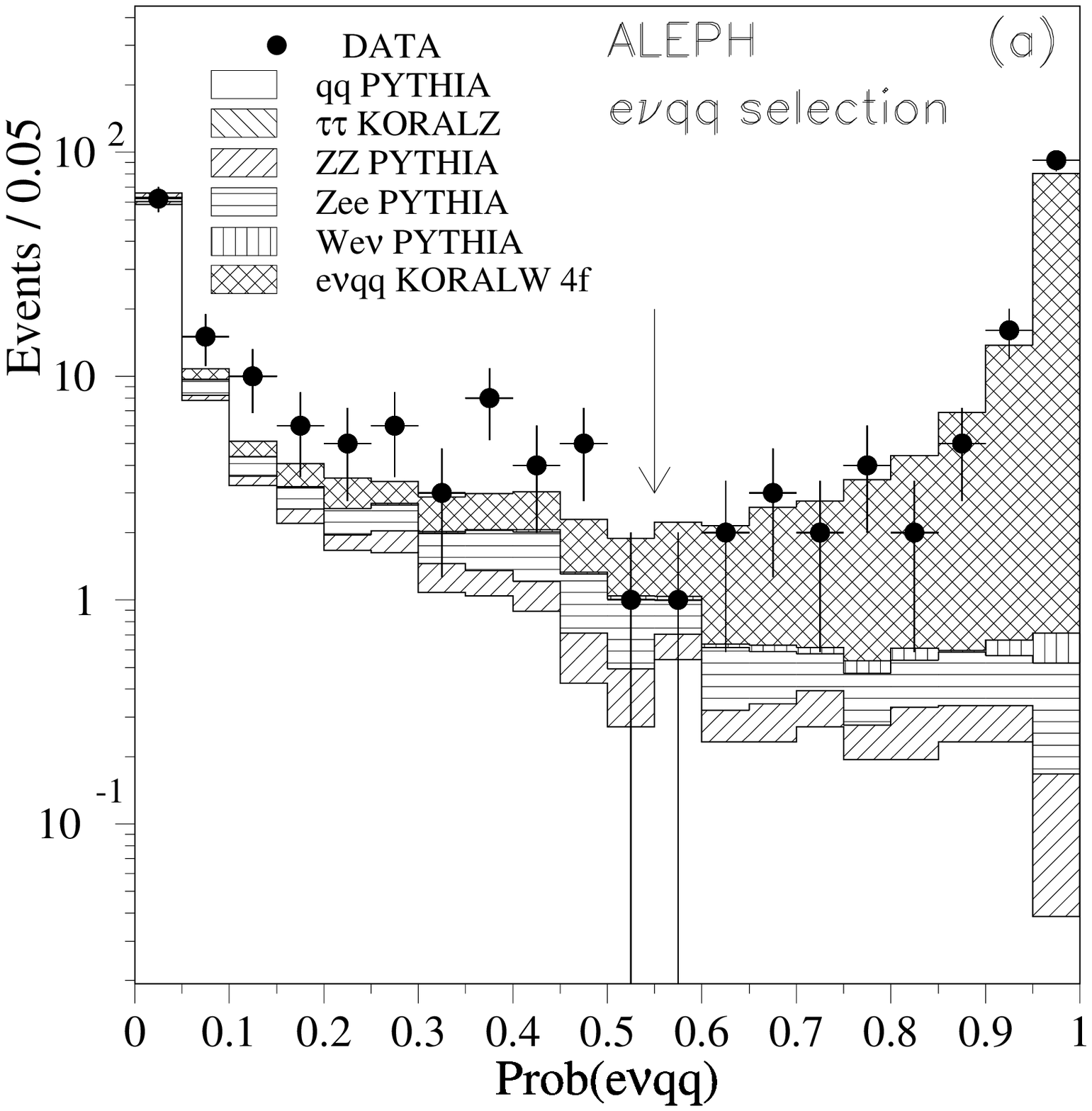,height=9cm}
    \epsfig{file=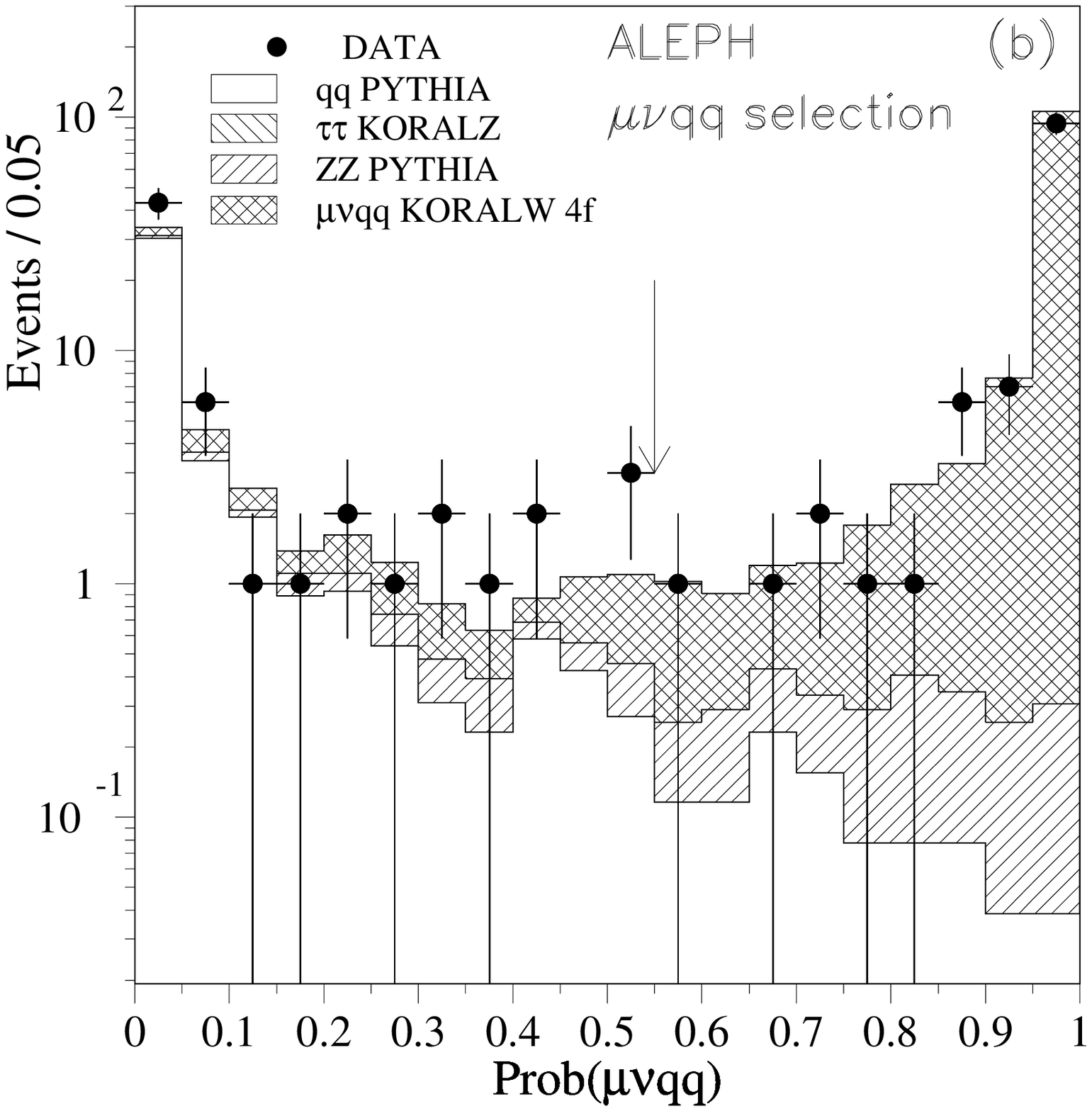,height=9cm}
    }
\vspace*{-0.5cm}
  \caption{\sl Probability distribution of preselected events for the 
    $\ev\qq$ (a) and the $\mu\nu \qq$ (b) selections.  
The arrows
show the cuts above which events are kept in the selection.}  
\label{Plvqq}
\end{figure}

The selection of $\ev \qq$ and 
$\mu\nu \qq$ events 
is basically unchanged with respect to 
the previous analysis~\cite{xsecA}. 
After the preselection of hadronic events 
based on charged energy and multiplicity, 
the charged track with the highest momentum
component antiparallel to the missing momentum
is chosen as the lepton candidate.
The same electron or muon identification criteria as 
for the fully leptonic channel are required for this lepton candidate 
track, as well as an energy of at least 15 GeV. 
The probability for an event to come from a signal process is determined
from the energy and isolation of the lepton and the total missing transverse 
momentum. 
Selected events are required to have a probability greater than
0.55, both for electron and muon candidates (Fig.~\ref{Plvqq}).

 The CC03 selection efficiencies are 81.5\%
for the $\ev\qq$ and 88.8\% for the 
$\mv\qq $ channel, with a total background of 158 fb;
127 electron and 113 muon events are selected in the data.

\boldmath
\subsubsection{$\PWW \ra \tau\nu \qq$ events}
\unboldmath
The selection of $\tau\nu \qq$ events is 
based on global event variables and is combined with a topological 
selection designed to identify the $\tau$ jet.
The common preselection is described in Ref.~\cite{xsecA}.
In the global analysis the acoplanarity is required
to be less than $175^{\circ}$, the 
energy in a wedge of half-angle $30^\circ$ 
centred on the missing momentum direction 
in the plane transverse to the beam axis
smaller than 12.5\%$\sqrt{s}$,
the estimated energy of the ``primary'' neutrino smaller than 65 \GEV, and
the visible mass smaller than 140 \GEVcc.
In the topological analysis the higher energy of 
the two quark jets must be 
smaller than 65 GeV and 
the acolinearity of the quark jets more than $110^{\circ}$.

In the data 238 events are selected by the combined analyses, 
of which 152 are already in the 
\ev\qq\ and \mv\qq\ samples.
The tau exclusive selection,
i.e.\ not counting events already selected 
by the electron and muon semileptonic selections, 
has an efficiency of 54.1\% 
and total background
of 114 fb, dominated by $\qqbar$ events. 

\subsubsection{Results}
The inclusive combined efficiencies for the three semileptonic
decay channels are 87.7\% for the electron channel,
92.3\% for the muon channel and 60.5\% for the tau channel, 
giving an 80.2\% 
overall efficiency for $\PWW\ra\lv\qq$, 
with a background of 270 fb (Table~\ref{tab:summ}).
The overall 4f-CC03 correction is +35 $\pm$ 36 fb.
A total of 322 events are selected in the data. 

The systematic uncertainties on the cross sections (Table~\ref{tabsys}) 
are dominated by the background normalisation, 
and by the uncertainty coming from the cut on energy around the 
beam for \tv\qq\ events. Since the selections of semileptonic events is 
basically unchanged, all systematic uncertainties on the 
signal efficiencies and on the backgrounds have been evaluated in the 
same way as for the lower energy measurements~\cite{xsecA}.

A similar fit as for fully leptonic events is used here with
the corresponding matrix of efficiencies and backgrounds
including the overlaps of the different $\lv\qq$ analyses.
The partial cross sections are then extracted from such a maximum
likelihood fit to the number of events in each selection, 
leading to
\begin{eqnarray*}
\sigma (\PWW\ra \ev\qq) & = &
 2.52 \pm 0.24 \stat 
\pm 0.06 \syst ~\pb, \\
\sigma  (\PWW\ra\mv\qq) & = &
 2.12 \pm 0.21 \stat
\pm 0.06 \syst ~\pb, \\
\sigma  (\PWW\ra\tv\qq) & = &
 2.15 \pm 0.31 \stat 
\pm 0.09 \syst ~\pb,
\end{eqnarray*}
where the systematic errors are obtained by varying all input
parameters of the fit
according to their uncertainties.

The total $\lv\qq$ cross section is extracted by means of the same
fit, under the  
assumption of lepton universality:

$$
\sigma  (\PWW\ra\ell\nu \qq) =
 6.81 \pm 0.40 \stat 
\pm 0.14 \syst~\pb.
$$

\boldmath
\subsection{$\PWW \ra \qq\qq$ events}
\unboldmath

The analysis of WW decays to four jets 
is updated from Ref.~\cite{xsecA} 
and consists of a simple preselection followed by a fit to the 
distribution of the output of a neural net (NN).

For each event, particles are clustered into four jets 
with the DURHAM algorithm~\cite{durham}. 
A first preselection cut rejects
events with a real Z and large undetected 
initial state radiation, requiring that the total
longitudinal momentum be smaller than 95\% of the difference between the 
visible mass and the mass of the Z. The sphericity must be larger than 0.03.
The value $Y_{34}$ 
of the jet resolution parameter 
where a four jet becomes a three jet event,
is required to be greater than 0.001. 
A further requirement that none of the four jets 
contains more than 95\% of electromagnetic energy rejects $\qqbar$ events 
with a visible initial state radiation photon.

\begin{figure}[htb]
\vspace{-1cm}
\begin{center}
\mbox{\epsfig{file=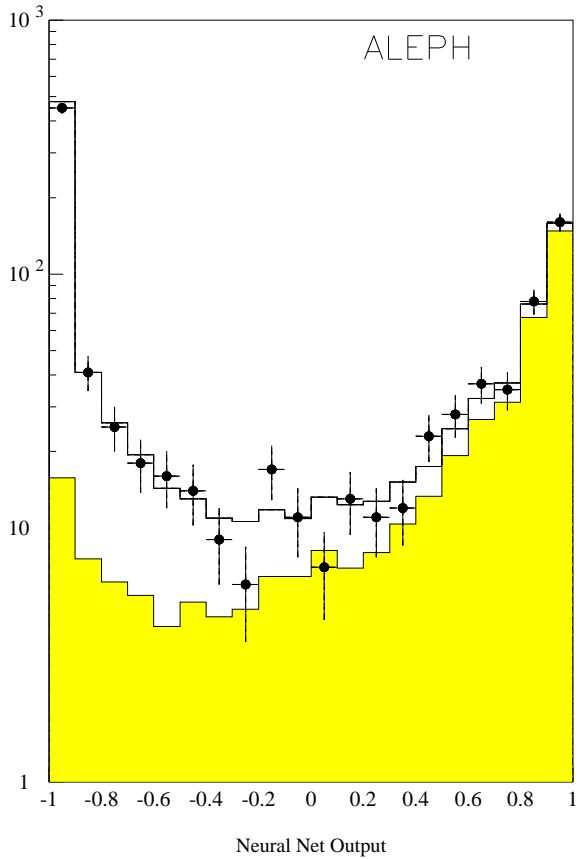,width=95mm}}
\end{center}
\vspace*{-0.5cm}
\caption{\sl Comparison of NN output distributions for data 
and Monte Carlo after the preselection. 
The points are the data and the open
histogram the total Monte Carlo prediction. The 
light shaded histogram shows the expected 
$\PWW \ra \qq\qq$ contribution.
\label{had:nn}}
\end{figure}

At this level of the selection, 1014 events are selected in the data while 
995.4 events are expected from all Standard Model processes. 
This preselection has an  
efficiency of 97\% on CC03 events and a purity of 40\%.

The input variables for the NN are described in Appendix A, and 
are related to the global event properties, 
the properties of jets, WW kinematics
and the b-tag probabilities for the four jets.
%
Figure~\ref{had:nn} gives the distribution of the NN output values for all 
backgrounds and signal Monte Carlo events compared with the data.

The number of signal events is extracted by means of a binned 
maximum likelihood fit 
to the distribution of the NN output for data events.
Only the normalisation of the MC signal is allowed to vary in the fit. 
The 4f-CC03 correction is $+10 \pm 8$fb.
 
The fit result is
$$
\sigma (\PWW\ra \qq\qq) =7.35 \pm 0.42\stat \pm 0.25\syst~\pb.
$$

Systematic errors are summarised in Table~\ref{tabsys}.
The uncertainty on the variables used for the 
event preselection cuts, and for the NN input 
(preselection and shape variables), has been assessed by
reweighting the events in the NN output according to bin by bin Data/MC 
differences in the distributions of such variables. 
Uncertainties with the WW generator have been evaluated comparing
{\tt KORALW} and {\tt EXCALIBUR} MC samples, {\tt EXCALIBUR} has 
also been used to evaluate the effects of colour reconnection.
Finally, for the QCD fragmentation uncertainty,
the {\tt HERWIG}~\cite{HERWIG} generator was used to produce 
different samples of \PWW~signal and \qqbar~background events.

\section{Total cross section}
\begin{figure}[htb]
\begin{center}
\mbox{
 \epsfig{file=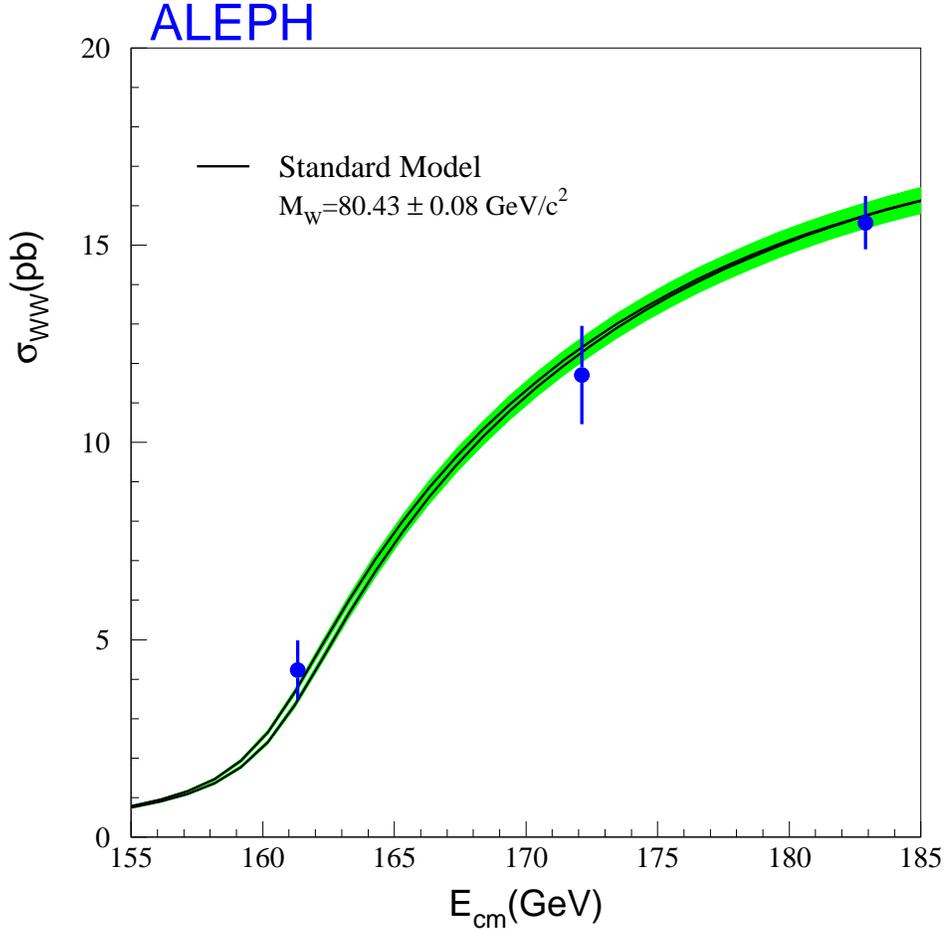,width=140mm}
}
\end{center}
\vspace*{-0.5cm}
\caption{\sl 
Measurements of the W-pair production 
cross section at three centre-of-mass energies, compared
to the Standard Model prediction from {\tt GENTLE}~\cite{GENTLE} 
for the world average value of the W mass~\cite{MWWA}. 
The two curves correspond
to the 80~\MEVcc~ error on~\PMW~, the shaded area around the
two curves represents the theoretical error ($\pm 2\%$) 
on the {\tt GENTLE} calculations.
\label{fig:wwxs}}
\end{figure}

The total cross section is obtained from a fit 
assuming the Standard Model branching ratios, 
the only unknown being the total cross section. 
The fit is applied to all data selected 
as described in the previous sections, and using the matrices of 
efficiencies and backgrounds 
of the various analyses, yielding
$$
\PSW =  15.57 \pm 0.62\stat \pm 0.29\syst ~\pb.
$$
Figure~\ref{fig:wwxs} shows the total cross section measured 
as a function of the c.m. energy.

\section{Branching ratios and $V_{\rm cs}$}
The same fit as for the total cross section 
is performed adding together the data samples collected at 
161, 172 and 183~\GEV~centre-of-mass energies. 

Without assuming lepton coupling universality, the six
unknowns are the three individual leptonic branching ratios 
and the three total 
cross sections at 161, 172 and 183~\GEV. 
The hadronic branching ratio is set to 
$1 - {\rm B}_{\mathrm e} - {\rm B}_{\mu} - {\rm B}_{\tau}$. 
The fitted leptonic branching ratios are
\begin{eqnarray*}
{\rm B}(\PW\ra\ev)&=&11.15\pm 0.85\stat\pm 0.24\syst\%,\\
{\rm B}(\PW\ra \mv)  &=& 10.06\pm 0.78\stat\pm 0.21\syst\%, \\
{\rm B}(\PW\ra \tv)  &=&  9.76\pm 1.01\stat\pm 0.33\syst\%,
\end{eqnarray*}
and are consistent with lepton universality 
and the Standard Model expectations.
Due to cross-contaminations in the identification of W decays to 
\tv~versus \ev~or \mv , the measured ${\rm B}(\PW\ra \tv)$ is
26\% anticorrelated with ${\rm B}(\PW\ra\ev)$ and
22\% anticorrelated with ${\rm B}(\PW\ra\mv)$.

If lepton universality is assumed 
(${\rm B}_{\mathrm e} = {\rm B}_{\mu} ={\rm B}_{\tau}=
 (1-{\rm B}_{\rm q})/3$), 
a fit to
${\rm B}(\PW\ra \qq)$ 
and the three total cross sections as unknowns 
yields the measurement of the hadronic branching ratio
$$
{\rm B}(\PW\ra \qq) = 68.93 \pm 1.21 \stat\pm 0.51  \syst\%.
$$

This result can be expressed in terms of 
the individual couplings of the
W to quark-antiquark pairs:
\begin{eqnarray}
   \frac{{\rm B}(\PW\ra \qq)}{1-{\rm B}(\PW\ra \qq)} & = &
 ( |\ckm{ud}|^2 + |\ckm{cd}|^2 + 
|\ckm{us}|^2 + |\ckm{cs}|^2 + |\ckm{ub}|^2 + |\ckm{cb}|^2 ) 
 (1+\alpha_{s}(\PMMW )/\pi)
\nonumber \\
\end{eqnarray}
Using the world average value of 
$\alpha_{s}$(\PMMZ ) evolved to \PMMW\,
together with the other measured CKM matrix 
elements~\cite{pdg98} allows a constraint 
on the least well measured CKM matrix element, 
$$
|V_{cs}| = 1.043 \pm 0.058 \stat\pm 0.026 \syst.
$$



\clearpage

\section{Invisible W width}

The cross section given in the previous sections are based on the assumption
that the W decays exclusively into the known quarks and leptons.
Scenarios of new physics have been advocated whereby the W could decay 
into a charged particle with momentum 
below detectability, $P<200~\MEVc$~\cite{kaliz}.
These decays would appear as invisible, in that events 
containing them would not be selected by the previously described 
selections.
A first consequence would be the appearance of events with only
one  visible W decay. 
A direct search for this scenario has been performed by ALEPH
and is described elsewhere~\cite{singlew}.

A second consequence would be a modification 
of the visible WW cross section as follows. 
The total width $\PGW$ of the W boson 
would be increased with respect to the Standard Model prediction for 
decays into known fermions, $\PGWv $,
by the corresponding invisible partial width,
$$
\PGW = \PGWv + \PGWi .
$$
The dominant effect would be a decrease of 
the visible cross section $\PSWv$ according to
$$
 \PSWv = (B^{\mathrm{vis}})^2 \PSW,
$$
where 
$$
 B^{\mathrm{vis}}=\frac{ \PGWv }{\PGWv + \PGWi } = 1-B^{\mathrm{invis}}.
$$
The total cross section $\PSW$ has a non-negligible, but small,
dependence upon a change of the total width, if one assumes that 
the couplings of the W
to known fermions are left unmodified.  
The above formulae have to be corrected for
mixed decays, in which one W boson decays
to a visible channel while the other W boson decay is undetected. 
These events would not  be completely missed by the standard
analyses, they have
an efficiency to be selected by  the $\tv\qq$ selection of
$22.4\%$ and $8.9\%$ at 172 and 183 GeV, respectively. 
Such events would not be detected by the other selections. 
This correction term is proportional to
$\PGWi / \PGWv $
and the above selection efficiency; it represents a change in the 
visible cross section at the per mil level.

The invisible width $\PGWi $ is extracted
by means of a fit to the measured cross sections 
at $\sqrt{s}$=161, 172 and 183 GeV,
taking into account common systematic errors 
in the theoretical prediction and 
in the event selection. The theoretical prediction of $\PSW$ is 
computed with GENTLE~\cite{GENTLE}, taking into account its residual 
sensitivity to a change in the total width.  
The possibility that the triple gauge boson couplings
could be anomalous is allowed, by fitting simultaneously 
$\Delta \kappa_\gamma, \lambda_\gamma$ and $\Delta g_1^Z$ 
within the constraints allowed by the analysis of angular distributions 
in the semileptonic channel~\cite{tgc172}.
The W mass is allowed to vary in the fit within the constraint imposed by the 
current world average value (not including
the W mass from the WW cross section measurements at LEP), 
${\PMW} = 80.43
\pm 0.08\:{\GEVcc}$~\cite{MWWA}.

In addition to the systematic errors on the cross section measurements, 
three more correlated systematic errors are considered: 
i)~a theoretical uncertainty of $\pm 2\%$ in the prediction of the total WW 
cross section, 
ii)~a theoretical uncertainty 
of $\pm 0.5\%$ in the Standard Model W width and
iii)~the beam energy uncertainty $\Delta E_b = \pm 25 $ MeV, 
where the last two sources of systematic uncertainty
have been found to have a negligible effect (below 1 MeV).
The different contributions to the total systematic error
are summarized in Table~\ref{FitallSyst}.

The  fit to the cross section data~\cite{xsecA} gives
\begin{eqnarray}
\PGWi &=& 
30\:^{+52}_{-48}\:{\rm (stat.)}\:\pm\:{33}\:{\rm (syst.)}\:{\rm MeV} \; (\chi^2/dof = 3.5/2), 
\nonumber\\
\PGWi &<& 139 \:{\rm MeV}~{\mathrm{~at~95\%~C.L.}}, 
\nonumber \\
B^{\mathrm{invis}}  &<& 6.5\%~{\mathrm{~at~95\%~C.L.}}
\nonumber
\end{eqnarray}
The corresponding  total W width  is
\begin{eqnarray}
\PGW &=& 
2.126\:^{+0.052}_{-0.048}\:{\rm (stat.)}\:\pm\:{0.035}\:{\rm (syst.)}\:{\rm GeV}. \nonumber
\end{eqnarray}

\begin{table}[htb]
\begin{center}
\caption[]  
{\sl
Summary of systematic uncertainties on the $\PGWi $ determination 
from the cross section measurement.}
\vspace{6pt}
\label{FitallSyst}
\begin{tabular}{|c||c|}
\hline
{Source} & $\Delta \PGWi \:[{\rm MeV}]$ \\
\hline\hline
{Cross section measurement}             & $\pm 24$ \\
{Theor. uncertainty on $\PSW$}   & $\pm 23$ \\
\hline\hline
{Total uncertainty}                      & $\pm 33$ \\
\hline
\end{tabular}
\end{center}
\end{table}

\section{Conclusions} 
Using an integrated luminosity of 57~$\ipb$ 
the W pair production cross section at $\sqrt{s}$ =  182.7 GeV has been 
measured in all decay channels. The determination of individual 
branching ratios has been performed. The total cross section is 
$15.57 \pm 0.62  \stat \pm 0.29 \syst ~\pb$. 
After adding the data taken at 161 and 172 GeV c.m. energy, 
the hadronic decay branching ratio is
$ 68.93 \pm 1.21 \stat \pm 0.51 \syst$\% 
which is used to determine the CKM matrix 
element $|V_{cs}| = 1.043 \pm 0.058\stat \pm 0.026\syst$.
These results are consistent with similar measurements 
performed at LEP at the same centre-of-mass energy~\cite{Olep}.

The agreement of the cross section with the Standard Model prediction 
allows a limit to be set on the W partial width into invisible decays:
$
\PGWi = 
30\:^{+52}_{-48}\:{\rm (stat.)}\:\pm\:{33}\:{\rm (syst.)}\:{\rm MeV} $, $
\PGWi < 139 \:{\rm MeV}~{\mathrm{~at~95\%~C.L.}}$

\section*{Acknowledgements}
It is a pleasure to congratulate our colleagues 
from the CERN accelerator divisions 
for the successful operation of LEP at 183 GeV.
We are indebted to the engineers and technicians in all our institutions for 
their contributions to the excellent performance of ALEPH. 
Those of us from non-member 
countries thank CERN for its hospitality.

\section*{Appendix A. Hadronic neural network input variables}

The neural network hadronic event selection uses 19 variables. These are based on 
global event properties, heavy quark flavour tagging, jet properties and
\PWW\ kinematics and are listed below together with their discriminating power.
The four jets are numbered in order of decreasing energy.

{\em Global event properties}

\begin{itemize}
 \item Fox Wolfram moment H0 (4.4\%)
 \item Fox Wolfram moment H2 (4.7\%)
 \item Fox Wolfram moment H4 (10.1\%)
 \item Sphericity (3.9\%)
 \item Missing Energy (4.4\%)
 \item Sum of the $p_t^2$ of all charged tracks to the beam axis (4.1\%)

{\em Heavy flavour tagging}

 \item Sum of the b-tag probabilities for the four jets (5.7\%)

The following jet related variables are determined from kinematically fitted jet 
momenta.

{\em Jet properties}

 \item Number of good charged tracks in jet1 (6.2\%)
 \item Maximum energy carried by one energy flow particle in Jet1 (3.8\%)
 \item Maximum energy carried by one energy flow particle in Jet2 (4.6\%)
 \item Maximum energy carried by one energy flow particle in Jet3 (4.7\%)
 \item Sum of the angles between the leading and remaining charged tracks in jet1 
(5.5\%)
 \item Sum of the angles between the leading and remaining charged tracks in jet2 
(3.6\%)

{\em \PWW kinematics}

 \item Sum of the cosines of the 6 angles between jets (8.7\%)
 \item Cosine of the angle between jet2 and jet3 (4.6\%)
 \item Energy of jet1 (8.1\%)
 \item Energy of jet2 (3.8\%)
 \item Momentum of jet4 (5.5\%)
 \item Asymmetry in momentum between jet2 and jet3 (3.8\%)
\end{itemize}
 

\end{document}